# CloudQTL:
# Evolving a Bioinformatics Application to the Cloud

**John Allen[14], David Scott[2], Malcolm Illingworth[2], Bartek Dobrzelecki[5], Davy Virdee[2], Steve Thorn[3], Sara Knott[4]**

[1]National e-Science Centre (NeSC), Edinburgh, EH8 9AA.
[2]EPCC, University of Edinburgh, JCMB, Edinburgh, EH9 3JZ,
[3]Information Systems, University of Edinburgh, Edinburgh, EH93JZ,
[4]Institute of Evolutionary Biology, University of Edinburgh, Edinburgh EH9 3JT,
[5]ABB Corporate Research Centre in Krakow, Poland; formerly of (2011) EPCC

John Allen (corresponding author)  john.allen@ed.ac.uk

**Abstract**

A timeline is presented which shows the stages involved in converting a bioinformatics software application from a set of standalone algorithms through to a simple web based tool then to a web based portal harnessing Grid technologies and on to its latest inception as a Cloud based bioinformatics web tool. The nature of the software is discussed together with a description of its development at various stages including a detailed account of the Cloud service. An outline of user results is also included.

**Introduction**

A quantitative trait is a phenotype or organism characteristic with continuous measurement such as product yield and quality in agricultural species or risk factors for disease in animal and human populations. It is usually complex in that it is influenced by the actions and interactions of many genes and environmental factors and geneticists are interested in identifying and understanding the role of the genes involved.

Quantitative trait locus mapping is a statistical modeling approach to identifying regions of the genome known as QTLs (Quantitative Trait Loci) that are involved in the control of the trait and is an essential tool for understanding the genetic basis of complex traits. It involves the use of molecular markers to follow inheritance of specific genome locations from parent to offspring and combines information from these with pedigree and trait records to look for associations between genotype and phenotype.

**1990s to 2005 – Standalone Application to the World Wide Web.**

Production and release of *QTL Express* [1], a user-friendly, web-accessible analysis tool, involved converting QTL mapping algorithms [2] initially written in Fortran into Java servlets. *QTL Express* allowed users to send data and receive output in series for simple QTL mapping analyses using moderately sized data of the order of kilobytes. It has seen wide use for the analysis of experimental data for QTLs, and it has received around 500 citations.

**2005-2010 - e-Science push - Grid Portal technologies**

The advent of microarray technologies that produce high-density multiple trait gene expression datasets and the availability of dense gene marker maps for thousands of individuals increased the dimensionality and complexity of QTL analyses requiring computationally intensive and more advanced QTL mapping tools. This led to a push for more computational power, a need to develop more complex QTL algorithms as well as the ability to accommodate more users using larger data sets of the order of megabytes as the QTL community grew.

*GridQTL* [3] & [4] provided an expanded and improved QTL analysis tool from *QTL Express* in a user friendly web portal environment, harnessing Grid technologies to deal with these increased computational demands and offering data persistence, parallel submission and retrieval of data with access via a user login to a personal data space for reviewing results. Work started in 2005 and involved collaboration with the Institute of Evolutionary Biology (IEB) at Edinburgh University, Roslin Institute, National e-Science Centre (NeSC), and EPCC (Edinburgh Parallel Computing Centre). The web portal was based on *GridSphere* [5] that acted as a container to the QTL algorithms that had evolved once more into JSR 168 compliant Java portlets [6]. The portal uses the power of the NGS [7], the Edinburgh Compute and Data Facility (ECDF) [8] & [9] and, for very large data sets HECToR [10], the UK's national high-performance computing service, in the computational Grid. Grid middleware from the *Globus Toolkit* [11], and *Enabling Grids for e-Science project, EGEE* [12] were used for job-submission and querying methods as well as for management tools for the authentication and authorisation processes involved in the use of the Grid resources. A typical view of the portal during an analysis run is shown in Figure 1.

*GridQTL* was first released in the autumn of 2006 and demonstrated at the UK e-Science All Hands conference of that year [13]. To date over 600 individual users have performed near to 100000 analyses in their QTL studies and are now using around half a cpu-year of computation time on our Grid per year. Around 50 users a month use *GridQTL* in every continent of the world; a map detailing the location of our users who have cited *GridQTL* is available from our website [4] and is shown in Figure 2. As of summer 2013 over 110 papers detailing QTL studies that have used and cited *GridQTL* have been published.

Examples of QTL Studies performed with *GridQTL* to date have included: resistance to disease in sheep [23]; growth in young cattle [24]; harvest traits in salmon [25]; domesticity studies in foxes [26]; obesity in mice [27]; wood quality of eucalyptus trees [28]; scale quality in crocodiles [29], airway obstructions in thoroughbred racehorses [30] and seed toxicity in oilseed crops [31], though more are available by exploring the links from the website. This short list of studies emphasises the wide variety of animal and plant studies that have been made with *GridQTL*.

**2010 and onwards – reaching for the Clouds.**

A further tranche of funding allowed for the inclusion of new QTL models in the portal as well as the investigation of Cloud computing. The *GridQTL* portal has so far given users access to the QTL algorithms and the computational resources free of charge; however, there is no way of sustaining this once the project funds run out.

Our view of Cloud Computing is in line with the view presented in [14]. Cloud Computing brings together Software as a Service (SaaS) and Utility Computing where Utility Computing is a service made available in a pay-as-you-go manner by the Cloud Provider. One can distinguish several classes of Utility Computing amongst the current Cloud computing offerings. The difference is based on the level of abstraction presented to the programmer wanting to access virtualised resources. For example the Google AppEngine [15] provides automatic scaling and load but enforces the programmer to use a predefined application structure and a fixed API; on the other side of the argument is Amazon's EC2 [16] which allows the author to control nearly the entire software stack There is also the middle ground represented by Microsoft's Azure platform [17] that supports general purpose computing but requires applications to be compiled to the specific runtime. *GridQTL* uses complex backend applications to perform calculations, and it was deemed to be too expensive to port these to new runtime environments. Only the fully virtualised model, similar to Amazon's EC2, was practical for moving the existing portal to Cloud infrastructure.

When developing *CloudQTL* we sought the Amazon route via the Open Source cloud middleware projects *Eucalyptus* [18] and *OpenStack* [19] middleware, both of which implement subsets of the EC2 API, using a prototype local test Cloud provided by the Edinburgh University ECDF Cloud; this would enable eventual Cloudbursting to similar Clouds implementing EC2 API. Development of *CloudQTL* has however been considered with other Cloud middleware in mind, e.g. *OpenNebula* [20], OCCI [21], so as not to tie the development to one specific access route to Cloud systems.

In order to make it easier to understand what needed to be done to integrate the *CloudQTL* code into the *GridQTL* Portal some of the workings of the existing code will be described next.

**Existing *GridQTL* Job Submission Mechanism**

From a user's perspective, running a current GridQTL job consists of the following stages:

- Upload QTL data to the portal
- Run initial processing of QTL data, with option to run locally or remotely.
- Retrieve results for a completed job
- Review processed results in portal GUI and select parameters for QTL analysis
- Submit main processing job, with option to run locally or remotely. Note that the same results may be used as input for multiple jobs.
- Retrieve results and review in portal GUI; download and display results in various formats.

The current *GridQTL* architecture consists of a set of servlets and a suite of job management scripts. The servlets provide a user management and presentation layer, and a job management queue. The job management queue can execute *GridQTL* jobs either locally or remotely. Remotely executed jobs are run by copying scripts and executables to a remote host, then managing the job via *Globus* commands in scripts on the portal server. This job submission mechanism has been designed to give to the user our goal of the *GridQTL* service – fast and reliable submission, analysis, retrieval and displaying of the users' QTL data and subsequent results.

As stated above *GridQTL* jobs fall into two categories - local analysis jobs to be run on the local server and remote jobs to be run on the Grid. A user can choose where the job is to be sent though *GridQTL* employs logic based on the size of the data set to automate this choice. *GridQTL* jobs are limited by memory to the maximum number of markers on a chromosome - roughly in proportion to the square of this value. Local jobs are converted to remote jobs when this number of markers exceeds 100 and the job is then directed onto the Grid. A *GridQTL* job with over 1200 maximum number of markers on a chromosome will need around 4GBytes of memory; the local ECDF Grid has approximately 1000 4GByte cores and the local *GridQTL* server has 8x4GBytes cores. For data with greater than 1200 maximum markers on a chromosome and up to 5000 of these markers (equivalent to 30 GBytes – a group of 8 cores) jobs can be run on the Grid by assembling groups of cores for increased RAM and running with one core (via *Oracle Grid Engine* [32] *qsub* command and appropriate options) or sent to HECToR though we seldom have to deal with such analyses (less than 0.1% of all analyses). Times of execution of jobs are affected by the number of markers in the data, sample size, phenotype information and type of QTL analysis and can vary from a few minutes up to tens of hours for the very large memory data sets mentioned above. On an average day around 40 analyses are run on the Grid and 20 sent locally. Sizes of data vary from Kbytes up to Gbytes but the local bandwidth together with an internal job scheduler (described below) easily cater for the transfer of data to and from the Grid with this rate of job submission. Peaks in job submission do occur –at twice the average rate pre-August and pre-December; on a daily basis, though *GridQTL* is used worldwide (see Figure 2), there is a one and half times the average rate of job submission during Western Europe working hours.

*GridQTL* employs an internal job scheduler to submit jobs to two queues. The first is a local job queue for our local server jobs and also for jobs that are being prepared to be submitted to the Grid; once a job is submitted to the Grid it is removed from the local queue and placed on the second queue – the remote job queue. This latter queue is in fact a list of Grid jobs and the scheduler periodically checks their status and downloads output once these remote jobs are finished or indeed cancelled or failed. The scheduler employs two thread pools that are used to place the local jobs in the queue for future execution. This is done because, in the case of a remote job, placing the job directly in a remote queue can take a significant amount of time (around 10 seconds for the larger size of data) and if a number of such jobs are submitted in quick succession time-outs can occur. The remote job queue employs another thread in the scheduler to execute the checking loop of the remote jobs' queue.

Usually the number of local jobs running is set to one (i.e. one local and one remote job targeted for the Grid). Depending on number of and size of jobs this number can change depending on load to a value that can be configured in the portal up to the number of cores-1 on our local server (currently eight cores). On the current *GridQTL* server up to seven small local jobs could then be running at once with seven jobs being prepared for submission to the Grid. This feature of *GridQTL* has proved itself to be robust with the rates and size of job submission discussed above. The various states of *GridQTL* jobs are highlighted in a viewer with colours to represent their state: queuing (pink), running (orange), completed (blue for local, green for grid jobs), and errors (red) (see fig 1).

**Details of CloudQTL Design**

The intention within the *CloudQTL* project was to add an additional service to the above by providing execution of *GridQTL* jobs on cloud-hosted virtual machines, without disrupting the existing facilities to execute jobs locally or on the grid.

The *CloudQTL* service consists of the following major components:

- Job Manager
- A database of jobs and virtual machines
- Queue Manager
- CloudQTL Instances
- Virtual Machines
- Virtual Machine Manager
- Cloud

How these components interact is sketched in the next subsection. Subsequent subsections describe, in outline, the separate components.

*Workflow*

Figure 3 sketches the way in which a QTL Cloud job is processed.

To submit a job, the portal first creates a multipart message which contains the type of the job and all of the input files required for the chosen *CloudQTL* application. The portal then creates an http request containing the message, and posts this to the Job Manager's REST API. The REST API is implemented using the *Jersey toolkit* [33], and the Job Manager is hosted in a suitable web service container such as *Apache Tomcat* [34]. On receiving the job request, the Job Manager creates a new entry in its job queue (Database). Periodically the Queue Manager, which runs in a thread contained in the Job Manager, checks the queue for new jobs. On encountering a new job, the Queue Manager requests a VirtualMachine instance (if one is available) from the Virtual Machine Manager (VMManager) running within it. The QueueManager then submits the job to the virtual machine's job service interface and sets the job's status in its own queue to "running". Once the job has finished executing on the virtual machine, the virtual machine posts the output from it back to the Job Manager via an http request. On receiving the output the Job Manager sets the job's status to "completed" and releases the VM so that it can execute another job.

The Job Manager, Queue Manager and VM Manager run in separate threads. All of them access the Database but only one thread is allowed to access the Database at any given time in order to prevent inconsistencies developing. This may appear at first sight to be an overcomplicated design but there are reasons for this structure. Firstly the job manager services HTTP requests and a new instance is created to service each request and is discarded afterwards. In contrast the Queue Manager and VM Manager continue to run throughout the lifetime of the *CloudQTL* service. They have different functions as their names suggest and having them run in separate threads allows them to carry out time consuming operations, such as starting a new virtual machine instance in the *QTL Cloud*, without interfering with the timely processing of HTTP requests by the Job Manager.

*The Job Manager*

The Job Manager accepts HTTP requests from the Portal and from Simple Job Services running on instances in the QTL Cloud. Most notably the Job Manager accepts requests from:

- the portal to place jobs in a queue for future execution,
- job services to accept results from completed jobs,
- the portal to transmit results from completed jobs to the portal.

*The Database*

The Database implements a job queue and also keeps track of the virtual machine instances that have been created in the QTL Cloud. The Database is manipulated by

- the Job Manager (which does such things as add jobs to and remove jobs from it)
- the Queue Manager (which does such things as request instances for the jobs to run on)
- the VM Manager (which does such things as furnish idle Virtual Machines).

The Database is implemented as a file accessed by database operations. It would have been simpler to implement the Database as a list but a database has been adopted with the idea that in the future its persistent nature can be used to create a system that would be more resilient in the case of failure of the *CloudQTL* system.

*The Queue Manager*

There is only ever one instance of the Queue Manager and it is present throughout the time that the *CloudQTL* service is provided. It examines the database periodically setting jobs running on idle virtual machine instances in the cloud. It also monitors the job queue and requests extra virtual machine instances if necessary from the Virtual Machine Manager. The Queue Manager does not send messages directly to a virtual machine instance in the cloud, instead it communicates with the associated instance of a Virtual Machine.

*CloudQTL Instances*

A virtual machine instance in the QTL Cloud is used to execute GridQTL jobs; it starts to run Tomcat whilst booting. This instance of Tomcat contains a web application which encapsulates the QTL application software so the instance can execute all types of GridQTL jobs; the application software may be run by sending HTTP requests to the web application. When a job that is running on an instance completes it pushes the results back to the Job Manager via an HTTP request.

The execution service is intended to run only one job at a time, and will reject submission requests while it is currently executing a job. The image is preconfigured to contain all of the GridQTL application jars, such that these do not need to be transmitted with the job submission request.

*Virtual Machines*

A VirtualMachine class represents an instance of a virtual machine running on a cloud host. The VirtualMachine instance is effectively a Java wrapper round an http client, and encapsulates the http commands necessary to communicate with the job execution service running on an instance in the QTL cloud. Each virtual machine instance in the QTL cloud is associated with an instance of the class VirtualMachine that knows how to send HTTP requests to the instance in the cloud.

*The Virtual Machine Manager*

The Virtual Machine Manager attempts to maintain a reasonable supply of virtual machine instances in the QTL cloud in order to allow the Queue Manager to run jobs. It is not desirable for jobs to be kept waiting for long periods nor for virtual machines to be idle for long periods. The Virtual Machine Manager attempts to balance these conflicting requirements. In order to do this it may either create new virtual machine instances or terminate idle ones. It does not do this directly however but via the class Cloud.

*The Cloud*

The class Cloud maintains a database describing the virtual machine instances that are present in the QTL Cloud. There are never any instances of this class. It will attempt to create new virtual machine instances or terminate existing instances when requested to do so by the Queue Manager.

The queue of submitted jobs and a list of available virtual machines are implemented as a db4o database which provides persistence for the job queue.

***CloudQTL* 2013**

*GridQTL* 3.3.x and *CloudQTL* 1.3.x were released in the summer of 2013 as a single portal (the name *GridQTL* being kept for historical reason) and featured the first version of a non-chargeable cloud system in use. The QTL cloud was finally built with middleware using the *AWS* Free-tier system. Another test version of *CloudQTL* was also produced that employed *OpenStack* to access the local ECDF Cloud;

successful analyses were completed on both systems.

**Results & Conclusions**

Expedience dictated the final choice of middleware. Edinburgh University's Information Systems built the local test cloud using middleware from *OpenStack* after having initially chosen *Eucalyptus*; the local expertise gained from this experience was crucial in building *CloudQTL*. For our production system using *Amazon AWS*, the ease of moving to this middleware from *OpenStack* proved very straightforward with only minor changes for exceptions handling and in handling messages generated when creating new instances in the Cloud; no changes had to be made to the overall design and the AWS Free-tier system currently gives enough free cpu time for current *CloudQTL* jobs. The authors would like to emphasis that the choice of middleware was not made with regard to one system being "better" or "easier" than the other. Such new and innovative software requires local help and experience and this was not at hand when using *OpenNebula* for this project. Investigations made with *OpenNebula* proved useful and the authors would like to note its ease of installation and extensive documentation from its release in the summer of 2013.

Some further work had to be considered before releasing *CloudQTL* with regard to security and encryption when transferring data between machines. The Job Manager and the VM Manager had to run on the same server as the existing portal server and were locked down to accept http requests from the local host. The one exception was the method for receiving data from a cloud image and a method was found to allow this communication to be secured via https. All data transmission between the Job Manager and the cloud virtual machines were via https.

To achieve the goal for a chargeable service that will give *CloudQTL* perpetuity a cost model accounting system based on EPCC's SAFE project [22] will be used for implementation with a simultaneous move to a chargeable tier of AWS. For this release the minimum charging period for the cloud service will have to be taken into account in particular when starting a new instance if a current instance could run several jobs within its current charging period.

**Acknowledgements**

*GridQTL* was funded by the United Kingdom Biotechnology and Biological Sciences Research Council (BEP2, BBS/B/1695X, 2006-2010) with further funding for *CloudQTL* via the United Kingdom Biotechnology and Biological Sciences Research Council (Bioinformatics and Biological Resources (BB/G022658/1, *GridQTL+*, 2010-2013)).

# Figures

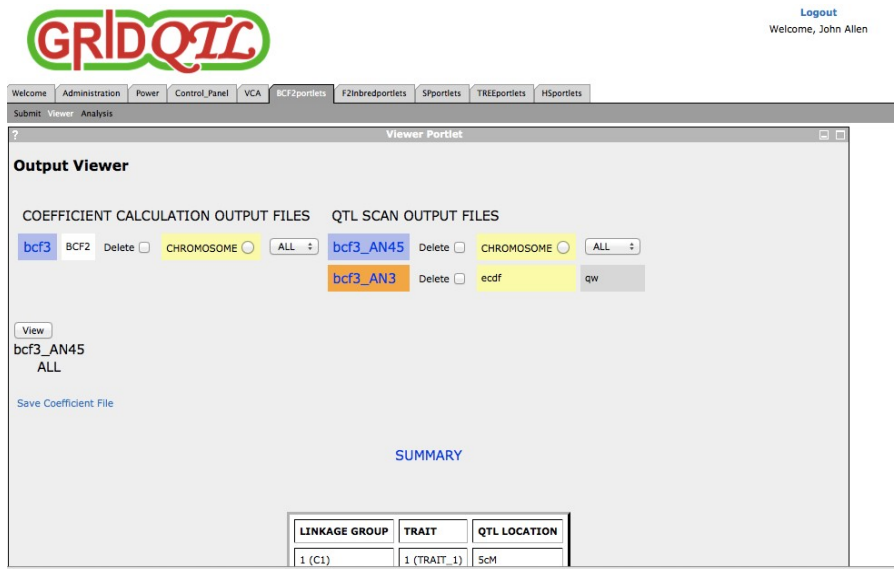

Figure 1 –Analysis Screen from the *GridQTL* Portal.

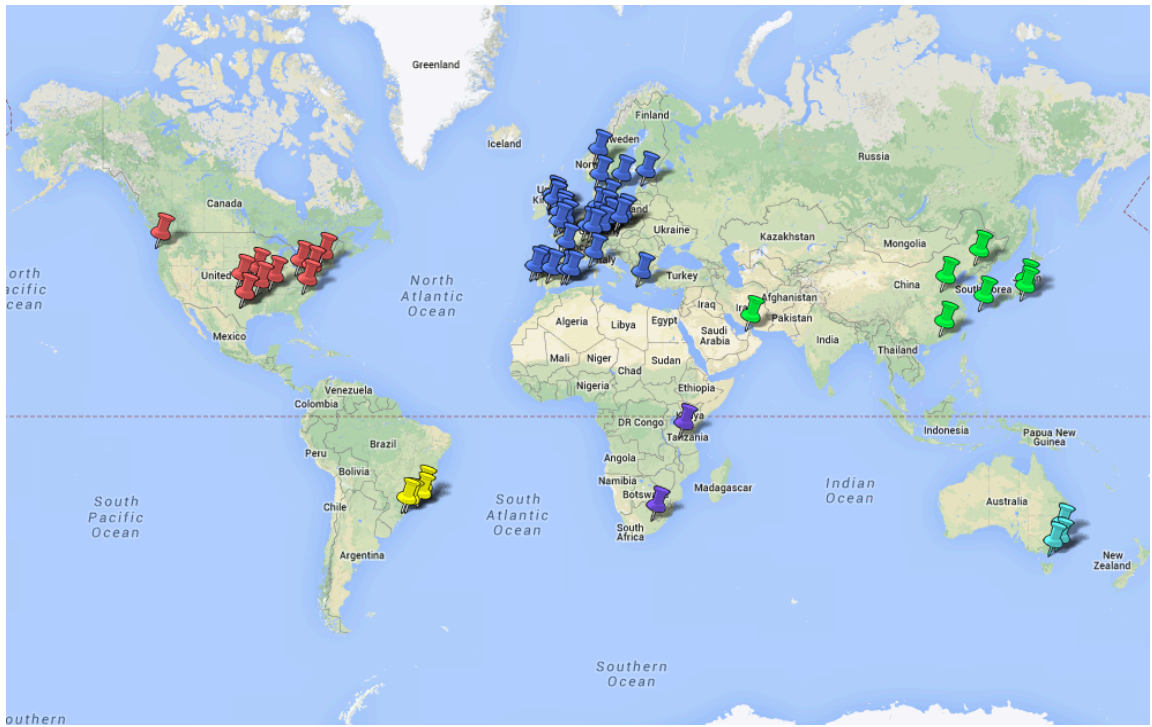

Figure 2 – GridQTL user citations by country.

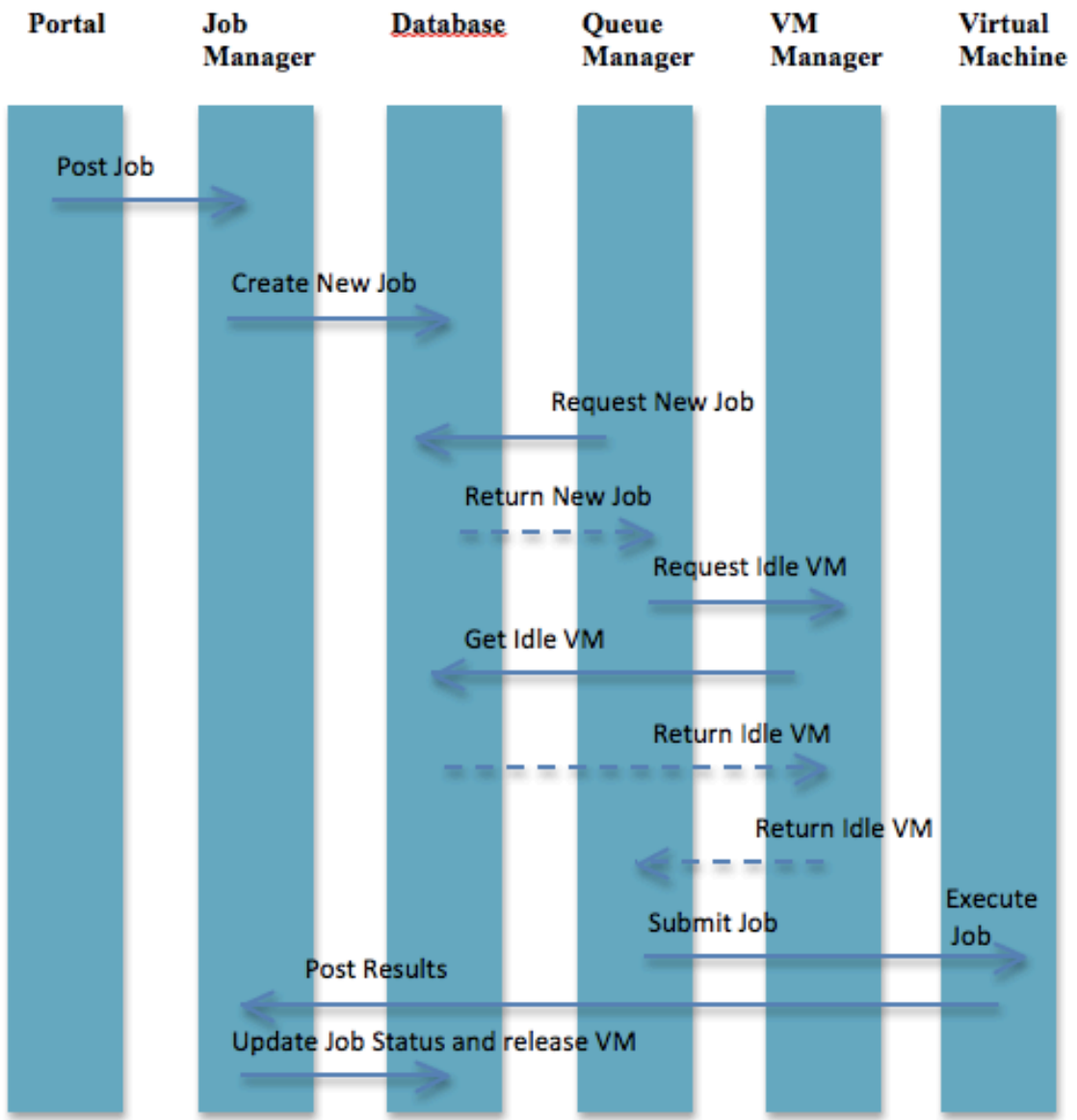

Figure 3 – CloudQTL WorkFlow